\documentclass[aps,preprint]{revtex4}%
\usepackage{amsfonts}
\usepackage{amsmath}
\usepackage{amssymb}
\usepackage{graphicx}%
\providecommand{\U}[1]{\protect\rule{.1in}{.1in}}
\begin{document}
\preprint{Physics of Plasmas}
\title{Nonlinear structures: explosive, soliton and shock in a quantum
electron-positron-ion magnetoplasma}
\author{R. Sabry$^{1,a}$, W. M. Moslem$^{1,b}$, F. Haas$^{1,c}$, S. Ali$^{2,d}$ and P.
K. Shukla$^{1,e}$}
\affiliation{$^{1}$Institut f\"{u}r Theoretische Physik IV, Fakult\"{a}t f\"{u}r Physik und
Astronomie, Ruhr-Universit\"{a}t Bochum, D-44780 Bochum, Germany}
\affiliation{$^{2}$National Center for Physics, Quaid-I-Azam University Campus, Islamabad, Pakistan}

\pacs{PACS }

\begin{abstract}
Theoretical and numerical studies are performed for the nonlinear structures
(explosive, solitons and shock) in quantum electron-positron-ion
magnetoplasmas. For this purpose, the reductive perturbation method is
employed to the quantum hydrodynamical equations and the Poisson equation,
obtaining extended quantum Zakharov-Kuznetsov equation. The latter has been
solved using the generalized expansion method to obtain a set of analytical
solutions, which reflect the possibility of the propagation of various
nonlinear structures. The relevance of the present investigation to the white
dwarfs is highlighted.

\end{abstract}
\received{14 October 2008}

\startpage{1}
\endpage{20}
\maketitle

\section{Introduction}

Numerous investigations \cite{1,2,3,4,5} relating to wave phenomena, have been
studied in dense quantum plasmas, are of fundamental importance for
understanding collective interactions in superdense astrophysical environments
\cite{6}, in high intense laser-solid density experiments \cite{7}, in
ultracold plasmas \cite{8}, in microplasmas \cite{9}, and in micro-electronic
devices \cite{10}\textrm{.} New characteristics of quantum plasma arise due to
the pressure law describing the fermionic behavior of the charged carriers,
quantum forces associated with the electron tunneling, as well as the Bohr
magnetization involving the electron 1/2 spin. The quantum Bohm potential
produces modifications in the dispersions of collective modes at quantum
scales. The latter are strongly effected by the plasma number densities and
Fermi temperatures. It is well-known that quantum mechanical effects become
relevant when the thermal de Broglie wavelength of the charged particles is
equal or larger than the average interparticle distance. In particular,
quantum behavior of the electrons reaches much easily due to less mass
compared to ions.

In recent years, many theoretical and numerical analysis \cite{11,12,13,14,15}
have been carried out to investigating the new features of plasmas with
quantum corrections by using both the Schr\"{o}dinger-Poisson and the
Wigner-Poisson systems. In this context, Manfredi \cite{11} reported different
approaches to model the collisionless electrostatic dense quantum plasmas.
Haas \textit{et al}. \cite{12} investigated the linear and nonlinear
properties of the quantum ion-acoustic (QIA) waves in dense quantum plasmas by
employing the quantum hydrodynamical (QHD) equations for inertialess electrons
and mobile ions. They examined that the quantum Bohm potential modifies the
linear wave dispersion and affects strongly the QIA solitary waves. Shukla and
Eliasson \cite{13} presented the numerical study of the dark solitons and
vortices in quantum electron plasmas. Moslem \textit{et al.} \cite{14}
investigated the quantum dust-acoustic double layers in a multi-species
quantum dusty plasma. It was found that both compressive and rarefactive
double layers can only exist for positively charged dust particles. Later, Ali
\textit{et al}. \cite{16} studied the QIA waves in a three-component plasma,
comprised of electrons, positrons, and ions. They employed the reductive
perturbation method and pseudo-potential approach for the small and arbitrary
amplitude nonlinear QIA waves, respectively. It was shown that the amplitude
and width are significantly altered due to the quantum statistics and quantum
tunneling effects. Misra \textit{et al.} \cite{17} considered the nonlinear
propagation of electron-acoustic waves in a nonplanar quantum plasma,
consisting of two groups of electrons: the inertial cold electrons and
inertialess hot electrons as well as the stationary ions. They obtained the
bright and dark solitons depending strongly upon the presence of cold electrons.

The laboratory and dense astrophysical quantum plasmas can be confined by an
external magnetic field. Therefore, the effect of the magnetic field has to
taken into account, especially for astrophysical observations (such as white
dwarfs, neutron stars, magnetars, etc.) where the high magnetic field plays an
important role in the formation and stability of the existing waves. Several
authors have considered the effect of magnetic field in different quantum
plasma models. For example, Haas \cite{18} introduced a three-dimensional QHD
model for dense magnetoplasmas and established the conditions for an
equilibrium\textbf{\ }in the ideal quantum magnetohydrodynamics (QMHD). Ali
\textit{et al. }\cite{19} employed the QMHD equations presenting a fully
nonlinear theory for ion-sound waves in a dense Fermi magnetoplasma. It was
revealed that only subsonic ion-sound solitary waves may exist. Shukla and
Stenflo \cite{20} derived the dispersive shear Alfv\'{e}n waves in a quantum
magnetoplasma, incorporating the strong electron and positron density
fluctuations. The shear Alfv\'{e}n modes acquire additional dispersion due to
quantum corrections. Later, Ali \textit{et al.} \cite{21} have been carried
out for the low-frequency electrostatic drift-like waves in a nonuniform
collisional quantum magnetoplasma. It was shown that the modes become unstable
and can cause cross-field anomalous ion-diffusion.

Three decades ago, Zakharov and Kuznetsov \cite{22} derived an equation for
nonlinear ion-acoustic waves in a magnetized plasma containing cold ions and
hot isothermal electrons. The Zakharov-Kuznetsov (ZK) equation has also been
derived for different physical systems and scenarios \cite{23,24}. Nonlinear
wave solution for ZK equation can produce an instability in a
three-dimensional system as discussed in Refs. \cite{25,26}. Moslem \textit{et
al}\textbf{. }\cite{27} extended the work for a three-dimensional nonlinear
ion-acoustic waves\ in a quantum magnetoplasma, highlighting the bending
instability of the solitary wave solution of the quantum ZK equation.
Recently, Masood and Mushtaq \cite{28} studied obliquely propagating
electron-acoustic waves in a two-electron population quantum magnetoplasma and
examining the effects of nonlinearity at quantum scales.

In the present paper, we shall investigate the possible nonlinear structures
(soliton, explosive and shock pulses) of the QIA waves in a collisionless
electron-positron-ion magnetoplasma using the QHD equations. By means of
computational investigations, we examine the effect of the positron
concentration, the quantum diffraction and the quantum statistical effects on
the profiles of the nonlinear excitations. The paper is organized as follows:
The basic equations governing the dynamics of the nonlinear QIA waves are
presented and the extended quantum ZK equation describing the system is
derived in Sec II. In Sections III and IV, we apply the generalized expansion
method to solve the extended quantum ZK equation. A set of analytical
solutions is obtained, and then used to investigate numerically the effect of
positrons and the quantum parameters on the nonlinear excitations. The results
are summarized in section V.

\section{Basic equations and derivation of the extended quantum ZK equation}

We consider a dense magnetoplasma whose constituents are the electrons,
positrons, and singly charged positive ions. The plasma is confined in an
external magnetic field $\mathbf{H}_{0}=H_{0}\widehat{\mathbf{z}},$ where
$\widehat{\mathbf{z}}$ is the unit vector along the $z-$axis and $H_{0}$ is
the strength of the magnetic field. We assume that the quantum plasma
satisfies the condition $T_{Fe,p}\gg T_{Fi}$, and obeys the electron/positron
pressure law $P_{e,p}=m$ $n_{e,p}^{3}V_{Fe,p}^{2}/3n_{e,p0}^{2}$, where
$V_{Fe,p}=\left(  2K_{B}T_{Fe,p}/M\right)  ^{1/2}$ is the electron/positron
Fermi thermal speed, $K_{B}$ is the Boltzmann constant, $T_{Fe,p}$ ($T_{Fi}$)
is the electron/positron (ion) Fermi temperature, $M$ is the electron and
positron mass, $n_{e,p}$ is the electron/positron number density, with the
equilibrium value $n_{e,p0}$. The nonlinear propagation of the QIA waves is
governed by the dimensionless hydrodynamics equations as%
\begin{equation}
\frac{\partial n_{i}}{\partial t}+\mathbf{\nabla.}\left(  n_{i}~\mathbf{u}%
_{i}\right)  =0, \tag{1}%
\end{equation}

\begin{equation}
\frac{\partial\mathbf{u}_{i}}{\partial t}+~\mathbf{u}_{i}\mathbf{.\nabla
u}_{i}=-\mathbf{\nabla}\phi+\mathbf{u}_{i}\times\widehat{\mathbf{z}}, \tag{2}%
\end{equation}%
\begin{equation}
\Omega\bigtriangledown^{2}\phi=n_{e}-n_{p}-n_{i}, \tag{3}%
\end{equation}%
\begin{equation}
n_{e}=\mu_{e}\left(  1+2\phi+H_{e}^{2}~\frac{\bigtriangledown^{2}\sqrt{n_{e}}%
}{\sqrt{n_{e}}}\right)  ^{\frac{1}{2}}, \tag{4}%
\end{equation}
and%
\begin{equation}
n_{p}=\mu_{p}\left(  1-2\sigma\phi+\sigma H_{e}^{2}~\frac{\bigtriangledown
^{2}\sqrt{n_{p}}}{\sqrt{n_{p}}}\right)  ^{\frac{1}{2}}, \tag{5}%
\end{equation}
where $n_{i}$, $\mathbf{u}_{i}$, and $\phi$ are the ion number density, the
ion fluid velocity, and the electrostatic potential, respectively. Since, the
ion mass is much larger than the electron/positron mass, one can ignore the
quantum effects of the ions in Eq. (2). The statistical and diffraction effect
for the system can be seen through the nondimensional parameters
$\sigma(=T_{Fe}/T_{Fp})$ \textbf{and} $H_{e}(=eH_{0}\hbar/2c\sqrt{M_{i}M}%
K_{B}T_{Fe})$\textbf{, respectively, where }$\hbar$\textbf{ is the Planck
constant divided by }$2\pi$\textbf{, }$M_{i}$\textbf{ }$(M)$\textbf{ is the
ion (electron/positron) mass, and }$c$\textbf{ is the speed of light in
vacuum. Here, }$\Omega(=\omega_{ci}/\omega_{pi})$\textbf{, where }$\omega
_{ci}$\textbf{ }$(=eH_{0}/m_{i}c)~$\textbf{and }$\omega_{pi}$\textbf{
(}$=\sqrt{4\pi e^{2}n_{i0}/M_{i}}$\textbf{) are the ion gyrofrequency and the
ion plasma frequency, respectively. }$n_{i0}$\textbf{ is the equilibrium ion
density.} Equations (4) and (5) reveal that the electrons and positrons do not
follow the Boltzmann law contrary to the classical plasma. The physical
quantities appearing in Eqs. (1)--(5) have been appropriately normalized:
$n_{e,i,p}\rightarrow n_{e,i,p}/n_{i0}$, $\mathbf{u}_{i}\rightarrow
\mathbf{u}_{i}/C_{s}$, $t\rightarrow t\omega_{ci}$, $\nabla\rightarrow
\nabla\rho_{s}$,~and $\phi\rightarrow e\phi/2K_{B}T_{Fe}$, where $\rho
_{s}(=C_{s}/\omega_{ci})$ is the ion-sound Fermi gyroradius and $C_{s}%
(=\sqrt{2K_{B}T_{Fe}/M_{i}})$ is the ion-sound Fermi speed.

Before going to the nonlinear developments, it is necessary to examine the
condition for neglecting the source term in the continuity equation due to
annihilation of plasma species. The details are given in the Appendix.

To investigate the propagation of QIA waves, we expand the dependent variables
$n_{e,i,p}$, $\mathbf{u}_{i}$, and $\phi$ about their equilibrium values in
power of $\epsilon,$%

\begin{align}
n_{i}  &  =1+\epsilon n_{i1}+\epsilon^{2}n_{i2}+\epsilon^{3}n_{i3}%
+...,\nonumber\\
n_{e,p}  &  =\mu_{e,p}+\epsilon n_{e,p1}+\epsilon^{2}n_{e,p2}+\epsilon
^{3}n_{e,p3}+...,\nonumber\\
u_{ix,y}  &  =\epsilon^{2}u_{ix,y1}+\epsilon^{3}u_{ix,y2}+\epsilon
^{4}u_{ix,y3}+...,\tag{6}\\
u_{iz}  &  =\epsilon u_{iz1}+\epsilon^{2}u_{iz2}+\epsilon^{3}u_{iz3}%
+...,\nonumber\\
\phi &  =\epsilon\phi_{1}+\epsilon^{2}\phi_{2}+\epsilon^{3}\phi_{3}%
+...,\nonumber
\end{align}
where $\epsilon$ is a keeping order parameter proportional to the amplitude of
the perturbation. Following the reductive perturbation method \cite{29}, we
express the independent variables into a moving frame in which the nonlinear
wave moves at a phase-speed of $\lambda$ (normalized with the ion-sound Fermi
speed $C_{s}$) as%

\begin{equation}
X=\epsilon x,\text{ \ \ \ \ }Y=\epsilon y,\text{ \ \ \ \ }Z=\epsilon\left(
z-\lambda t\right)  \text{ \ and \ \ \ }T=\epsilon^{3}t. \tag{7}%
\end{equation}
The neutrality condition at equilibrium reads $\mu_{e}=1+~\mu_{p}$, where
$\mu_{e}=n_{e0}/n_{i0}$ and $\mu_{p}=n_{p0}/n_{i0}$. Subistituting (6) and (7)
into Eqs. (1)--(5), we obtain the lowest-order in $\epsilon$ as%

\begin{align}
n_{i1}  &  =\frac{1}{\lambda^{2}}\phi_{1},\text{ \ \ \ }u_{ix1}=-\frac
{\partial\phi_{1}}{\partial Y},\text{ \ }\nonumber\\
\text{\ }u_{iy1}  &  =\frac{\partial\phi_{1}}{\partial X},\text{
\ \ \ \ \ }u_{iz1}=\frac{1}{\lambda}\phi_{1},\tag{8}\\
\text{\ \ \ \ \ \ \ \ \ \ \ \ \ }n_{e1}  &  =\mu_{e}\phi_{1},\text{
\ \ \ \ \ }n_{p1}=-\sigma\mu_{p}\phi_{1},\nonumber
\end{align}
along with the phase speed rule%
\begin{equation}
\lambda=\left(  \frac{1}{1+~\mu_{p}(1+\sigma)}\right)  ^{1/2}. \tag{9}%
\end{equation}
It is clear here that the phase speed $\lambda$ of the QIA waves is affected
by the quantum statistical effect and by the positron concentration $\mu_{p}$.
To the next-order in $\epsilon$, we have%

\begin{align}
n_{i2}  &  =\frac{4}{3\lambda^{4}}\phi_{1}^{2}+\frac{1}{\lambda^{2}}\phi
_{2},\text{\ \ ~\ \ ~}u_{ix2}=\lambda\frac{\partial^{2}\phi_{1}}{\partial
X\partial Z}-\frac{\partial\phi_{2}}{\partial Y},\nonumber\\
u_{iy2}  &  =\lambda\frac{\partial^{2}\phi_{1}}{\partial Y\partial Z}%
+\frac{\partial\phi_{2}}{\partial X},\text{ \ \ \ \ \ \ }u_{iz2}=\frac
{1}{2\lambda^{3}}\phi_{1}^{2}+\frac{1}{\lambda}\phi_{2},\tag{10}\\
n_{e2}  &  =-\frac{\mu_{e}}{2}~\left(  \phi_{1}^{2}-2\phi_{2}\right)
,\text{\ ~\ \ \ \ }n_{p2}=-\frac{\sigma\mu_{p}}{2}\left(  \sigma\phi_{1}%
^{2}+2\phi_{2}\right)  ,\nonumber
\end{align}
while the Poisson equation gives%
\begin{equation}
Q\phi_{1}^{2}=0, \tag{11}%
\end{equation}
where%

\[
Q=\frac{\left[  \left(  \sigma^{2}-1\right)  \mu_{p}\lambda^{4}-\lambda
^{4}-3\right]  }{2\lambda^{4}}.
\]
Since $\phi_{1}\neq0$, therefore $Q$ should be at least of the order of
$\epsilon.$ Therefore, $Q\phi_{1}^{2}$ becomes of the order of $\epsilon^{3}$;
so it should be included in the next order of the Poisson equation. The
next-order in $\epsilon$ gives a system of equations. Solving this system with
the aid of Eqs. (8)-(10), we finally obtain the extended quantum ZK equation
as%
\begin{equation}
\frac{\partial\varphi}{\partial T}+\left(  A~\varphi+B~\varphi^{2}\right)
\frac{\partial\varphi}{\partial Z}+C\frac{\partial^{3}\varphi}{\partial Z^{3}%
}+D\frac{\partial}{\partial Z}\left(  \frac{\partial^{2}}{\partial X^{2}%
}+\frac{\partial^{2}}{\partial Y^{2}}\right)  \varphi=0, \tag{12}%
\end{equation}
where~we have replaced $\phi_{1}$ by $\varphi$ for simplicity. The nonlinear
and dispersion coefficients are given, as
\begin{align*}
A  &  =\frac{\lambda^{4}+3-\left(  \sigma^{2}-1\right)  \mu_{p}\lambda^{4}%
}{2\lambda},\\
B  &  =-\frac{3\left(  \left(  \sigma^{3}+1\right)  \mu_{p}\lambda^{6}%
+\lambda^{6}-5\right)  }{4\lambda^{3}},\\
C  &  =\frac{1}{8}\lambda^{3}\left(  4\Omega-\left(  \mu_{p}+1\right)
H_{e}^{2}-\sigma^{2}H_{e}^{2}\mu_{p}\right)  ,\\
D  &  =C+\frac{1}{2}\lambda^{3}%
\end{align*}
The extended quantum ZK equation (12) constitutes the final outcome of this
model. The anticipated balance between dispersion and nonlinearity (which
contain the quantum mechanical effects) within the extended quantum ZK
equation may give rise to different nonlinear structures. Some of these
solutions will recover in the next section.

\section{Exact solutions of the extended quantum ZK equation}

To obtain the possible analytical solutions of Eq. (12), we assume that
\begin{equation}
\xi=L_{X}X+L_{Y}Y+L_{Z}Z-\vartheta T, \tag{13}%
\end{equation}
where $L_{X}$, $L_{Y}$ and $L_{Z}$ are the direction cosines and $\vartheta$
is the QIA wave speed to be determined later. Using (13) into (12), we obtain%

\begin{equation}
-\vartheta\varphi^{\prime}+A_{0}\varphi~\varphi^{\prime}+B_{0}\varphi
^{2}\varphi^{\prime}+\gamma\varphi^{\prime\prime\prime}=0, \tag{14}%
\end{equation}
where $A_{0}=AL_{Z}$, $B_{0}=BL_{Z}$ and $\gamma=CL_{Z}^{3}+DL_{Z}\left(
L_{X}^{2}+L_{Y}^{2}\right)  $. According to the generalized expansion method
\cite{30} the solution of Eq. (14) can represent by
\begin{equation}
\varphi=a_{0}+a_{1}\omega, \tag{15}%
\end{equation}
with%

\begin{equation}
\frac{d\omega}{d\xi}=k\left(  c_{0}+c_{1}\omega+c_{2}\omega^{2}+c_{3}%
\omega^{3}+c_{4}\omega^{4}\right)  ^{1/2}, \tag{16}%
\end{equation}
where $a_{0},a_{1},c_{0},c_{1},c_{2},c_{3}$ and $c_{4}$ are arbitrary
constants to be determined later and $k=\pm1$. Substituting Eq. (15) into Eq.
(14) and making use of Eq. (16), we obtain a polynomial equation in $\omega$.
Equating the coefficients of different powers of $\omega$, we obtain an
overdetermined system of algebraic equations which can be solved with the help
of symbolic manipulation package \textit{Mathematica} to give three Jacobi
elliptic doubly periodic type solutions as
\begin{multline}
\varphi=-\frac{A_{0}}{2B_{0}}+k\sqrt{\frac{6~\gamma~c_{2}~m^{2}}{B_{0}~\left(
2m^{2}-1\right)  }}~\operatorname{cn}\left(  \sqrt{\frac{~c_{2}~}{~\left(
2m^{2}-1\right)  }}\xi\right)  ,\text{ }\nonumber\\
\text{with }c_{0}=-\frac{c_{2}^{2}m^{2}\left(  1-m^{2}\right)  }{c_{4}\left(
2m^{2}-1\right)  ^{2}},~~c_{2}>0,\text{ }c_{4}<0, \tag{17}%
\end{multline}%
\begin{multline}
\varphi=-\frac{A_{0}}{2B_{0}}+k\sqrt{\frac{6~\gamma~c_{2}~}{B_{0}~\left(
2-m^{2}\right)  }}~\operatorname{dn}\left(  \sqrt{\frac{~c_{2}~}{~\left(
2-m^{2}\right)  }}\xi\right)  ,\nonumber\\
\text{ with }c_{0}=\frac{c_{2}^{2}\left(  1-m^{2}\right)  }{c_{4}\left(
2-m^{2}\right)  ^{2}},~~c_{2}>0,\text{ }c_{4}<0, \tag{18}%
\end{multline}
and%
\begin{multline}
\varphi=-\frac{A_{0}}{2B_{0}}+k\sqrt{\frac{6~\gamma~c_{2}~m^{2}}{B_{0}~\left(
m^{2}+1\right)  }}~\operatorname{sn}\left(  \sqrt{\frac{-~c_{2}~}{~\left(
m^{2}+1\right)  }}\xi\right)  ,\nonumber\\
\text{ with }c_{0}=\frac{c_{2}^{2}m^{2}}{c_{4}\left(  m^{2}+1\right)  ^{2}%
},~c_{2}<0,\text{ }c_{4}>0, \tag{19}%
\end{multline}
where $m$ is a modulus of the Jacobian elliptic function and $c_{1}=c_{3}=0$.
As $m\rightarrow1$, the Jacobi doubly periodic solutions (17) and (18)
degenerate to the bell-shapped solitary wave
\begin{equation}
\varphi=-\frac{A_{0}}{2B_{0}}+k\sqrt{\frac{6~\gamma~c_{2}}{B_{0}}%
}~\operatorname{sech}\left(  \sqrt{c_{2}}\xi\right)  ,\text{ } \tag{20}%
\end{equation}
where the arbitrary constant $c_{0}$ vanishes. Again, as $m\rightarrow1$ the
solution (19) can degenerate to the kink-type wave solution%

\begin{equation}
\varphi=-\frac{A_{0}}{2B_{0}}+k\sqrt{\frac{3~\gamma~c_{2}}{B_{0}}}%
~\tanh\left(  \sqrt{\frac{-c_{2}}{2}}\xi\right)  , \tag{21}%
\end{equation}
where $c_{0}=c_{2}^{2}/4c_{4}.$ In the solutions (17)-(21), the QIA wave speed
$\vartheta=\frac{1}{2}\left(  -A_{0}^{2}/2B_{0}+2\gamma c_{2}\right)  $ where
$c_{2}\neq A_{0}^{2}/4\gamma B_{0}$.

Furthermore, the generalized expansion method provides us with further
analytical solutions of the extended quantum ZK equation (12) as
\begin{multline}
\varphi=-\frac{2~c_{2}~}{c_{3}+k~\sqrt{c_{3}^{2}-4c_{2}c_{4}}\cosh\left(
2\sqrt{c_{2}}\xi\right)  ~},\text{ }\nonumber\\
\text{with }c_{0}=c_{1}=0,\text{ }c_{2}=\frac{\vartheta}{\gamma},\text{ }%
c_{3}=-\frac{A_{0}}{3\gamma},\text{ }c_{4}=-\frac{B_{0}}{6\gamma}, \tag{22}%
\end{multline}
and%
\begin{equation}
\varphi=-\frac{A_{0}}{2B_{0}}\left[  1+k~\coth\left(  \sqrt{\frac{-A_{0}^{2}%
}{24\gamma B_{0}}}\xi\right)  \right]  ,\text{
\ \ \ \ \ \ \ \ \ \ \ \ \ \ \ \ \ with \ \ }\vartheta=-\frac{A_{0}^{2}}%
{6B_{0}}\text{ \ \ \ and \ \ }B_{0}<0. \tag{23}%
\end{equation}

\section{Parametric Analysis for White Dwarfs}

It is clear that the propagation speed of the QIA wave is modified by the
effect of the quantum statistical effect $\sigma$ and by the presence of
positrons $\mu_{p}$. As $\sigma$ and $\mu_{p}$ increase, the propagation speed
of the QIA wave will decrease. The dependence of the nonlinear structures
amplitude and width on the equilibrium positron number density ($\mu_{p})$ and
quantum effects $\sigma$ and $H_{e}$ is more perplex. First, it is important
to note that changing $\mu_{p}$ leads to a change in the phase-speed
$(\lambda)$ of the QIA waves $\left[  \text{see Eq. (9)}\right]  ,$ as well as
the electron concentration (via the charge-neutrality condition $\mu_{e}=1+$
$\mu_{p}$). Since the electron (positron) Fermi temperature depends upon the
equilibrium electron (positron) number density, it can also be affected by
$\mu_{p}$ through the charge-neutrality condition. As a result, the quantum
statistical ($\sigma$) and diffraction ($H_{e}$) effects will vary with the
positron concentration $\mu_{p}$.

Based upon the above findings, we shall now investigate the effects of the
relevant physical quantities, namely the positron concentration $\mu_{p}$ on
the profiles of the QIA nonlinear structures. We have used, as a starting
point, a typical set of plasma parameter values for white dwarfs \cite{11} (in
the absence of positrons; $\mu_{p}=0$), namely: $n_{e0}=4\times10^{28}$
$\mathrm{cm}^{-3},$ $T_{Fe}=4.9\times10^{8}$ $\mathrm{K},$ $\omega
_{ci}=1.88\times10^{16}$ $\mathrm{s}^{-1}$ and $\omega_{pi}=2.63\times10^{17}$
$\mathrm{s}^{-1}$. However, once the positrons species density is determined,
the values of $T_{Fe},$ $\lambda$ and $H_{e}$ are subsequently computed,
according to the above formulae, which also determine $A,$ $B,$ $C$ and $D.$
In the plots, we shall change the positrons concentration, which leads to
recalculate all the physical parameters again. Obviously, by varying the
positron concentration, we simultaneously modify all the parameter values used
in the plots below.

\subsection{Solitary and Explosive/Blowup Excitations}

It may be appropriate to point out that the analytical solutions in Sec. III
have been obtained for different arbitrary constants $k,$ $c_{0},...c_{4}.$
One of them is the localized solution (22), which is a bell-shapped solitary
wave solution. Recall that the arbitrary constant $k$ can be either $+1$ or
$-1$. For $k=-1$, a positive solitary pulse can propagate and for $k=+1,$ a
negative solitary pulse exist. Note that we have executed the negative
solitary pulse since it is not physically correct in the the present model.
Figure 2 depicts the QIA solitary pulse for different values of positron
concentration $\mu_{p}$, which now determines $T_{Fe,p}$ (and the ratio
$\sigma$) through the charge-neutrality condition. It is found that the
amplitude of the soliton pulse decreases by increasing $\mu_{p},$ resulting an
increase (decrease) of the electron Fermi temperature $T_{Fe}$ (quantum
diffraction effect $H_{e}$). Physically, the increase of $T_{Fe}$ leads to an
increase of the electron Fermi energy (viz. $K_{B}T_{Fe}=E_{Fe}\equiv
(\hbar^{2}/2m)(3\pi^{2}n_{e0})^{2/3}$), and as a result the ion Fermi energy
should decrease to conserve the energy law. The decrease of the ion Fermi
energy decreasing the nonlinearity of the system and hence the height of the
soliton pulse shrinks.

It may be interesting to note that for certain values of plasma parameters the
solitary pulse convert to an explosive/blowup excitation as shown in Fig. 3.
The blowup excitation indicates that an instability in the system can produce
due to the effect of the nonlinearity (which in our case depends on the
positron concentration $\mu_{p}$ and the quantum statistical effects $\sigma
$). On the other hand, the magnitude of some quantities (e.g. temperature,
pressure, density, etc.) leads to prejudice the balance between the dispersion
and the nonlinearity \cite{31}. Therefore, the amplitude may increase to very
high values, which gives rise to increasing the electric potential and then
accelerate the moving particles.

It is important to notice that Eq. (23) is an explosive/blowup solution, i.e.
the potential $\varphi$ infinitely grows at a finite point (for any fixed
$X,Y,$ $Z\rightarrow X_{0},Y_{0},$ $Z_{0}$), there exist an $\xi_{0}$ at which
the solution (23) blowup and thereby we regard the latter as an explosive
solution as depicted in Fig. 4.

\subsection{Shock/Double Layer Excitation}

For the shock/double layer solution \cite{32}, the boundary condition
$\varphi(\xi)\rightarrow0$ at $\xi\rightarrow\infty$ must satisfy. Applying
the last boundary condition into Eq. (21), we obtain the double layer solution
as%
\begin{equation}
\varphi=\varphi_{m}\left[  1+\tanh\left(  W_{D}\xi\right)  \right]  ,\text{ }
\tag{24}%
\end{equation}
where the amplitude of the double layers is $\varphi_{m}=-A_{0}/2B_{0},$ the
width is $W_{D}=\sqrt{-24\gamma B_{0}/A_{0}^{2}}.$ Here $\vartheta(=-A_{0}%
^{2}/6B_{0})$ is the shock wave speed. Notice that $B_{0}<0$ has to be
fulfilled, in order for making the width $W_{D}$ real. The numerical analysis
in Fig. 5, however, shows that for small positron concentration $\mu_{p}$ the
dominant situation corresponds to $B_{0}<0,$ so the double layers may exist.
For large positron concentration $\mu_{p},$ double layers cannot occur, since
$B_{0}>0.$ Typically, we have used the plasma density value for white dwarf
\cite{11} via $n_{i0}=2\times10^{32}$ cm$^{-3}$ and assume that $L_{z}=0.2,$
which leads to the fact that for negative $B_{0}$ (i.e., formation of double
layers) the positron concentration $n_{p0}$ must less than $1.43308\times
10^{31}$ cm$^{-3}.$ Also, it noted that the narrow range of $\mu_{p}$
[corresponding to $B_{0}<0$] will not change the ion gyrofrequency $\Omega.$
Generally speaking, one can also note from Eq. (24) that the nature of the
double layer depends on the sign of $A_{0},$ i.e. for $A_{0}>0$ a positive
double layer exists (viz $\varphi_{m}>0$), whereas for $A_{0}<0$ we would have
a negative double layer ($\varphi_{m}<0$). For white dwarf plasma parameters,
it is found that $A_{0}$ is usually greater than zero and then only positive
double layers can exist.

Equation (24) describes the double layer potential, which has a well-know
profile (cf. Fig. 6). This profile may change due to vary of physical
parameters. The dependence of double layer characteristics on the positron
concentration $\mu_{p}$ [which determines $T_{Fe,p},$ $H_{e}$ and $\sigma$
through the charge-neutrality condition] is depicted in Fig. 7. It is obvious
that an increase in the positron concentration $\mu_{p}$ shrinks the double
layers width but the amplitude increases by increasing $\mu_{p}.$

\textbf{It important to note here that in Ref. \cite{33}, the soliton
excitation in \textsl{e-p-i} magnetoplasma was investigated but the present
work investigates soliton, shock and explosive excitations in \textsl{e-p-i}
magnetoplasma. Therefore, the present model studies another two nonlinear
structures, which did not discuss in Ref. \cite{33}. Also, in Ref. \cite{27},
the authors used the extended Conte's truncation method to obtain the
solitary, explosive, and periodic solutions of the QZK equation. Note that
this method gives solitary and explosive excitations described by equation
(25) and periodic excitation described by equation (26). Thus, the extended
Conte's truncation method cannot predict the shock formation, which may arise
due to the presence of weakly double layers. In the present work, we have used
generalized expansion method. The later succeeded to describe soliton,
explosive, as well as shock excitations. Therefore, the present method can be
considered as a powerful tool to deal with more general nonlinear partial
differential equations. }

\section{Summary}

To summarize, we have presented the properties of the nonlinear structures QIA
waves in a very dense Fermi plasma, composed of the electrons, positrons and
positive ions. By employing the reductive perturbation method, an extended
quantum ZK equation is derived. The latter has been solved using the
generalized expansion method to obtain a set of analytical solutions, which
reflects the possibility of propagation of various nonlinear structures (viz.
explosive, soliton and shock pulses). We have numerically examined the effects
of the positron concentration (which changes the quantum statistics and
quantum diffraction parameters through the charge-neutrality condition) on the
electrostatic potential excitations, by varying relevant physical parameters.
It is found that the amplitudes and widths of the nonlinear structures are
significantly affected by the positron concentration, quantum statistical, and
quantum tunneling effects. Also, for certain plasma parameters the solitary
pulse transforms to blowup pulse. Finally\textbf{,} we stress that this
investigation should be useful for understanding the features of the nonlinear
structures QIA waves in an electron-positron-ion plasma, such as those in the
superdense white dwarfs and in the intense laser-solid matter interaction experiments.

{\Large Appendix: The necessary condition to neglect the annihilation process}

To neglect the annihilation process, the following inequality must satisfy%

\begin{equation}
\frac{1}{\omega_{pe}}<<T_{ann}, \tag{A1}%
\end{equation}
where $(1/\omega_{pe})$ is the electron plasma period and $T_{ann}$ is the
annihilation time. For nonrelativistic plasma, the time of annihilation reads
\cite{34}%

\begin{equation}
T_{ann}=\frac{4}{3\sigma_{T}n_{e}c}\left[  \frac{\Theta}{1+6\Theta}\right]  ,
\tag{A2}%
\end{equation}
where $\sigma_{T}$ ($=6.65\times10^{-25}$ cm$^{2}$) is the cross section and
$\Theta(=K_{B}T/mc^{2})$ is the temperature range, which satisfy the
inequality \cite{34}%

\begin{equation}
\alpha^{2}<\Theta<1, \tag{A3}%
\end{equation}
where $\alpha$ ($=7.2974\times10^{-3}$) is the Fine-structure constant.
Equation (A3) can be rewritten in terms of temperature as%

\begin{equation}
3\times10^{5}<T\text{ }(\text{K})<5.9\times10^{9} \tag{A4}%
\end{equation}
Inserting Eq. (A2) into (A1), we obtain%

\begin{equation}
\Theta>2.66\times10^{-19}n^{1/2}. \tag{A5}%
\end{equation}
Using Eq. (A3) and (A5), one can calculate the range of the density where the
annihilation can be ignored%

\begin{equation}
3.9\times10^{28}<n_{e}\text{ }\left(  \text{cm}^{-3}\right)  <1.4\times
10^{37}. \tag{A6}%
\end{equation}
The quantum effects become important for certain values of density ($n_{e,p}$)
and temperature ($T_{e,p}$). The quantum condition $n_{e,p}\lambda_{B}%
^{3}\geqslant1$ specifies the temperature-density relation, where the quantum
effects become important as%

\begin{equation}
T_{e,p}\leqslant3.2\times10^{-11}n_{e,p}^{2/3}. \tag{A7}%
\end{equation}
Using Eq. (A6) with (A7), one can calculate the range of temperature in
quantum plasma as%

\begin{equation}
3.6\times10^{8}<T\,_{e,p}\text{ }(\text{K})<1.8\times10^{14}. \tag{A8}%
\end{equation}
It is clear that the range for neglecting annihilation is well satisfied for
white dwarf [see Ref. \cite{11}]. Therefore, the present model can be
applicable to the dense white dwarf.

\begin{acknowledgments}
R.S. acknowledges the financial support from the Egyptian Government under the
Postdoctoral Research Program. The work of W.M.M. was partially supported by
Ruhr-Universit\"{a}t Bochum through the Framework of the HGF Impulse and
Networking Fund/ FZ-J\"{u}lich (Project Number: S080200W). W.M.M. also thanks
Professor R. Schlickeiser for his hospitality. F.H. thanks the financial
support from the Alexander von Humboldt Stiftung (Bonn, Germany).
\end{acknowledgments}

\begin{center}
\_\_\_\_\_\_\_\_\_\_\_\_\_\_\_\_\_\_\_\_\_\_\_\_\_\_\_\_\_\_\_\_\_\_\_\_\_\_\_\_\_\_\_\_\_
\end{center}

(a) Also at: Theoretical Physics Group, Department of Physics, Faculty of
Science, Mansoura University, Damietta Branch, New Damietta 34517, Egypt.
Electronic mail: refaatsabry@mans.edu.eg and sabryphys@yahoo.com

(b) Present address: Department of Physics, Faculty of Science-Port Said, Suez
Canal University, Egypt. Electronic mail: wmmoslem@hotmail.com and wmm@tp4.rub.de

(c) Also at: Universidade do Vale do Rio dos Sinos-UNISINOS, Av. Unisinos,
950, 93022-000 Sao Leopoldo RS, Brazil. Electronic mail: ferhaas@tp4.rub.de

(d) Electronic mail: shahid\_gc@yahoo.com

(e) Also at: Department of Physics, Ume\aa ~University, SE-90187 Ume\aa ,
Sweden. Electronic mail: ps@tp4.rub.de

\bigskip

\newpage

{\LARGE Figure Captions}

\textbf{Figure 1 (color online):}

Three-dimensional profile of the solitary pulse [given by Eq. (22)].
A\ positive solitary pulse for $k=-1$, $\mu_{p}=0.8304$, $\sigma=1.693$,
$\Omega=0.05$, $H_{e}=0.03,$ $T=0$, $Y=0.1$, $L_{x}=0.01$, and $L_{z}=0.1$.

\bigskip

\textbf{Figure 2 (color online):}

Two-dimensional profile of the solitary pulse [given by Eq. (22)].
A\thinspace\ positive solitary pulse for $k=-1$. For curve A, $\mu_{p}=0.5$,
$\sigma=2.08$, $\Omega=0.01$, and $H_{e}=0.0075$, for curve B, $\mu_{p}=0.75$,
$\sigma=1.75$, $\Omega=0.0102$, $\ $and $H_{e}=0.0068$ and for curve C,
$\mu_{p}=1$, $\sigma=1.587$, $\Omega=0.010291$, $\ $and $H_{e}=0.00624$. Also,
we have used $T=0$, $X=Y=0.1$, $L_{x}=0.01$, and $L_{z}=0.1$.

\bigskip

\textbf{Figure 3 (color online):}

Three-dimensional profile of the explosive/blowup pulse [given by Eq. (22)].
A\thinspace\ positive explosive pulse for $k=-1$, $\mu_{p}=0.6$, $\sigma=1.9$,
$\Omega=0.0102$, $H_{e}=0.0072,$ $T=0$, $Y=0.1$, $L_{x}=0.01$, and $L_{z}=0.1$.

\textbf{Figure 4:}

Three-dimensional profile of the explosive/blowup pulse [given by Eq. (23)],
for $\mu_{p}=0.0525$, $\sigma=7.37$, $\Omega=0.0257$, $H_{e}=0.0081,$ $T=0$,
$Y=0.1$, $L_{x}=0.01$, and $L_{z}=0.2$.

\textbf{Figure 5:}

The nonlinear coefficient $B_{0}$ is depicted against the positron density
$n_{p0}$ for $n_{i0}=2\times10^{32}$ cm$^{-3}$ and $L_{Z}=0.2.$ Recall that
for $n_{p0}<1.43308\times10^{31}$ cm$^{-3}$ the nonlinear coefficient
$B_{0}<0$ and then a shock pulse can propagate.

\textbf{Figure 6:}

Three-dimensional profile of the shock pulse [given by Eq. (24)], for $\mu
_{p}=0.05$, $\sigma=7.6$, $\Omega=0.03$, $H_{e}=0.008,$ $T=0$, $Y=0.1$,
$L_{x}=0.01$, and $L_{z}=0.2$.

\textbf{Figure 7:}

Two-dimensional profile of the shock pulse [given by Eq. (24)]. For curve A,
$\mu_{p}=0.05$, $\sigma=7.6$, $\Omega=0.03$, $H_{e}=0.008$ and for curve B,
$\mu_{p}=0.052$, $\sigma=7.37$, $\Omega=0.03$, and $H_{e}=0.0081.$ Here,
$T=0$, $X=Y=0.1$, $L_{x}=0.1$, and $L_{z}=0.2$. Recall that the narrow range
of $\mu_{p}$ will not affect on the ion-gyrofrequency $\Omega.$

\bigskip

\bigskip

\newpage
\end{document}